\documentclass[12pt,tightenlines,
    onecolumn,
    preprintnumbers,
    amsmath,
    amssymb,
    prd,
    showpacs,
]{revtex4}
\usepackage{color}
\usepackage{graphicx}%
\usepackage{amsmath}
\usepackage[normalem,normalbf]{ulem}
\usepackage{amssymb}
\usepackage{bm}
\usepackage{enumerate}

\newcommand{\nn}{\nonumber}

\begin{document}
\title{Black string and velocity frame dragging}

\author{Jungjai Lee}
\email{jjlee@daejin.ac.kr}
\affiliation{ Department of Physics, Daejin University,
Pocheon, 487-711, Korea.
}%

\author{Hyeong-Chan Kim}
\email{hckim@phya.yonsei.ac.kr}
\affiliation{Department of Physics, Yonsei University,
Seoul 120-749, Republic of Korea.
}%

\date{\today}%
\bigskip
\begin{abstract}
\bigskip

We investigate velocity frame dragging with the boosted
Schwarzschild black string solution and the boosted Kaluza-Klein
bubble solution, in which a translational symmetry along the
boosted $z$-coordinate is implemented.
The velocity frame dragging effect can be nullified by the
motion of an observer using the boost symmetry along the $z-$coordinate if it is not compact.
However, in spacetime with the compact $z-$coordinate, we show that the effect cannot be removed since the compactification breaks the global Lorentz boost symmetry.
As a result, the comoving velocity is dependent on $r$ and the momentum parameter along the $z-$coordinate becomes an observer independent characteristic quantity of the black string and bubble solutions.
The dragging induces a spherical ergo-region around the black
string.

\end{abstract}
\maketitle

After the discovery of Schwarzschild black hole solution~\cite{M}
in general relativity, there has been an enormous increase of
interest in black objects such as several kinds of black hole,
black string, and black $p-$brane.
The black holes in (3+1)-dimensional Einstein-Maxwell theory are
classified by a few parameters (so called hairs) such as mass $M$,
angular momentum $J$, and charge $Q$.
By the three parameters, the solutions are classified into four
specific families, the Schwarzschild metric~\cite{M}, the
Reissner-Nordstr\"{o}m metric~\cite{Q1,Q2}, the Kerr
metric~\cite{J}, and the Kerr-Neumann metric~\cite{n}.
In the presence of matters other than electromagnetic field,
several solutions are possible with different hairs, non-abelian
hairs~\cite{nA_1,nA_2,nA_3,nA_4,nA_5}, dilatonic hair~\cite{dilatonic_1,dilatonic_2,dilatonic_3,dilatonic_4}, quantum
hair~\cite{quantum_1, quantum_2}, and so forth.
In addition to these, there are another kind of solutions with
different asymptotic spacetime structure such as the BTZ
black hole~\cite{btz_1, btz_2}.

These parameters plays a role in changing the structure of black
objects.
The mass $M$ develops event horizon and the charge $Q$ allows a
black object to interact with gauge field.
The angular momentum ($J\neq 0$) induces a strong frame dragging and
develops an ergo-region around its event horizon.
The rotational frame dragging effect was first derived from the theory of general relativity in 1918 by Lense and Thirring, and is also known as the Lense-Thirring effect~\cite{LT_1, LT_2, LT_3}.
The rotational frame dragging has been measured in recent
experiments~\cite{rot_1,rot_2,rot_3,rot_4, rot_5}, which shows
that the general relativity is correct.

At first look, it seems that the momentum parameter $P$ does
not play a role in changing the spacetime structure and can be
gauged away by the simple coordinate transform like the boost.
Contrary to this naive expectation, in this letter, we discuss
that the boosted spacetime with non-zero momentum $P$ is not
equivalent to the static one if the boosted coordinate is compact.
Let us argue this using the boosted black
string~\cite{chodos,hovdebo} with the compact fifth $z-$coordinate
with $z=0$ being identified with $z=L$.
An analogy of twin paradox helps us understand the situation.
Let one of the twin circumnavigate along the compact $z$-coordinate with
constant velocity and meet the other staying at $z=0$.
If the relativity of velocity holds, the twin paradox reemerges since
none of the twin experience acceleration.
For this paradox being resolved, the relativity of velocity should be
broken for compact coordinate.
This manifestly shows that the rigid Lorentz boost along the compact
coordinate is not a symmetry of the spacetime any more.
In fact, the rigid Lorentz boost spoils the space-like feature of the
transformed space coordinate.
Therefore, the rigid Lorentz boost along the $z-$coordinate is excluded in the diffeomorphism group of the compact spacetime any more.

Actually, the velocity frame dragging effect induced by $P$ is not
well known and controversial.
If the momentum $P$ do the role in changing the spacetime
structure, the velocity frame dragging effect will be observable
in gravitating objects, such as black objects~\cite{inst_0, inst_1, inst_2, inst_3, inst_4, inst_5, inst_6} in string theory and other higher dimensional gravity theory~\cite{randal_1, randal_2,hawking}.
If there is a velocity frame dragging in these objects, the
momentum flow ($P\neq 0$) can induce an ergo-region around its
event horizon to alter the spacetime structure.
In particular, the black string with a compact coordinate or the
Kaluza-Klein bubble solution may exhibit this velocity frame
dragging effect.
This field is not touched at all so far.
In this paper, we investigate the frame dragging effect in these solutions by comparing it with that in the Kerr black hole.
We study how the frame dragging effect changes when the spacetime
has a compact fifth coordinate.

The metric of the Schwarzschild black string solution in (4+1) dimensions is
\begin{eqnarray} \label{metric:Schwarzschild}
ds^2 = -\left(1-\frac{2M}{r}\right)dt^2+dz^2+
    \frac{1}{1-2M/r}\, dr^2+
    r^2 \left(d\theta^2+\sin^2\theta d\phi^2\right).
\end{eqnarray}
The Kaluza-Klein bubble solution~\cite{harmark} is given by the double-Wick rotation $t\rightarrow iz$ and $z\rightarrow i t$ of Eq.~(\ref{metric:Schwarzschild}):
\begin{eqnarray} \label{metric:bubble}
ds^2 = -dt^2+\left(1-\frac{2M}{r}\right)dz^2+
    \frac{1}{1-2M/r}\, dr^2+
    r^2 \left(d\theta^2+\sin^2\theta d\phi^2\right).
\end{eqnarray}
We see that there is a minimal 2-sphere of radius $2M$ located at $r = 2M$ for this bubble solution.
To avoid a conical singularity at $r=2M$ we need that $z$ is a periodic coordinate with period $L = 8\pi M$.
Clearly, the solution asymptotes to $\mathcal{M}^4 \times S^1$ for $r\rightarrow \infty$.

Let us check the velocity frame dragging effect in these two solutions~(\ref{metric:Schwarzschild}) and (\ref{metric:bubble}).
Consider an observer moving with the velocity $v=\tanh \xi$ with the
coordinates transform,
\begin{eqnarray} \label{boost:asym}
\left(\begin{tabular}{c}
    ${t'}^0$\\ ${z'}^4$ \\\end{tabular}\right)
    &=&\left(\begin{tabular}{c}
    $\cosh \xi~~ -\sinh\xi$\\ $-\sinh\xi~~
        \cosh\xi$ \\ \end{tabular}\right)
\left(\begin{tabular}{c}
    $t$\\ $z$ \\\end{tabular}\right),
\end{eqnarray}
with respect to the static metrics~(\ref{metric:Schwarzschild}) and (\ref{metric:bubble}).  Then, a stationary metric can be obtained to be
\begin{eqnarray} \label{Sol}
ds^2 =g_{\mu\nu}dx^\mu dx^\nu&=&-F(r)dt^2 +2 X(r) dt dz + H(r) dz^2\\
&&+\frac{1}{1-2M/r} dr^2+r^2\left(d\theta^2+\sin^2\theta d\phi^2\right)
    ,\nn
\end{eqnarray}
where the functions $F(r)$, $H(r)$, and $X(r)$ are,
\begin{eqnarray*} \label{FHX:Sch}
F(r)&=&1- \frac{2  M\cosh^2\xi}{r} ,\quad
H(r)= 1+\frac{2  M\sinh^2\xi}{r}
   \quad
X(r) = \frac{M\sinh2\xi}{r},
\end{eqnarray*}
for the boosted Schwarzschild solution and
\begin{eqnarray*} \label{FHX:bubble}
F(r)&=&1+\frac{2  M\sinh^2\xi}{r} ,\quad
H(r)= 1-\frac{2  M\cosh^2\xi}{r}\quad
X(r) = -\frac{M\sinh2\xi}{r},
\end{eqnarray*}
for the Kaluza-Klein bubble solution, respectively.
The mass, tension, and the momentum flow along $z-$ coordinate, $M_{ADM}$, $\tau$, and $P$ are
\begin{eqnarray} \label{Maj:Kcq}
M_{ADM}&=& \left(1\pm \frac{\cosh 2\xi}{3}\right)
    \frac{3M}{4}\,, \\
\tau &=& \left(1\mp\frac{\cosh2\xi}{3}\right)
    \frac{3 M}{4} \, ,\nn \\
P &= & \pm\frac{M\sinh2\xi}{4} ,\nn
\end{eqnarray}
where the upper/down sign in Eq.~(\ref{Maj:Kcq}) is for the boosted Schwarzschild/bubble solution.
In the boosted Schwarzschild solution, the velocity frame dragging is induced by the momentum $P$.
However, an observer related with the coordinates transform~(\ref{boost:asym}) does not see any dragging effect and see a static metric.
This is one of main difference of the momentum driven metric from that of the well-known Kerr solution.

Note that the bubble solution with $L=8\pi M$ becomes naked (conical) singular at $r=2M$ if it is the boosted~\cite{chodos}.
Therefore, the boost along the $z$-coordinate alters the physical properties of the $r=2M$ surface.
Instead, we could choose the compactification length
$L_{boost}= 8\pi M \cosh\xi$ for the boosted bubble solution to avoid the naked singularity.
We can conclude that a given static bubble solution is inequivalent to the boosted bubble solution.

Before we study the black string solution with compact $z-$coordinate,
let us consider the Kerr blackhole with metric,
\begin{eqnarray} \label{eq:Kerr}
ds^2&=&-\left(1-\frac{2 M r}{\rho^2}\right) dt^2
    +\frac{\rho^2}{\Lambda^2} dr^2+\rho^2 d\theta^2  \\
    &+&\left(r^2+a^2+
    \frac{2M r\alpha^2}{\rho^2} \sin^2\theta\right)\sin^2\theta d\phi^2
    -\frac{4M r\alpha}{\rho^2} \sin^2\theta \,dt d\phi \,, \nn
\end{eqnarray}
where $\alpha= J/M$, $\rho^2=r^2+\alpha^2\cos^2\theta$, and $\Lambda^2= r^2-2M r+ \alpha^2$.
In the non-relativistic limit where $M$ goes to zero, the Kerr metric becomes the orthogonal metric for oblate spheroidal coordinate,
\begin{eqnarray*}
ds^2 = -dt^2+\frac{\rho^2}{r^2+\alpha^2} dr^2+\rho^2 d\theta^2+(r^2+
    \alpha^2)\sin^2\theta d\phi^2 .
\end{eqnarray*}
We may re-write the Kerr metric~(\ref{eq:Kerr}) in the following form:
\begin{eqnarray*}
ds^2&=&-\left(g_{tt}-\frac{g_{t\phi}^2}{g_{\phi\phi}}\right) dt^2
    +g_{rr} dr^2+g_{\theta\theta} d\theta^2 +g_{\phi\phi}
    \left(d\phi +\frac{g_{t\phi}}{g_{\phi\phi}}dt \right)^2 \,.
\end{eqnarray*}
This metric is equivalent to a co-rotating reference frame that is rotating with angular speed $\Omega =\omega(r,\theta)$ that depends on both the radius $r$ and the colatitude $\theta$,
\begin{eqnarray}\label{eq:omega}
\omega(r,\theta)=-\frac{g_{t\phi}}{g_{\phi\phi}}=
    \frac{2M\alpha r}{\rho^2(r^2+\alpha^2)+2M \alpha^2 r \sin^2\theta} .
\end{eqnarray}
Thus, an inertial reference frame is drawn by the rotating central mass, the frame dragging.
An extreme version of frame dragging occurs within the ergosphere of a rotating black hole. The ergosphere of the Kerr metric is bounded by two surfaces on which it appears to be (coordinate) singular. The inner surface corresponds to a spherical event horizon at $r_H=M+\sqrt{M^2-\alpha^2}$, where the purely radial component $g_{rr}$ of the metric goes to infinity. The outer surface is the stationary limit,
\begin{eqnarray*} \label{eq:r:ergo}
r_{stationary}= M+\sqrt{M^2-\alpha^2 \cos^2\theta} ,
\end{eqnarray*}
which touches the event horizon at the poles of the rotation axis, where the colatitude $\theta$ equals $0$ or $\pi$.

What is the difference of the frame dragging effect in the present
stationary solution~(\ref{Sol}) from that in the Kerr blackhole?
We have asymptotic observer at $r=\infty$ in black string metric in
which frame dragging may exist.
However, the frame dragging is the same at all points of the spacetime
if it is observed in the coordinates of the asymptotic observer.
Therefore, the dragging effect can be nullified by the motion of the asymptotic observer with velocity $v=\tanh\xi$.

An interesting question is whether we can identify the presence of
frame dragging effect or not, if we are restricted to a circle with fixed radius and polar angle in Kerr spacetime.
To begin with, let us review a well known thought experiment, which
defines a {\it locally non-rotating observer}, in the Kerr spacetime (See Exercise 33.3 of Ref.~\cite{misner}).
Place a rigid, circular mirror (``ring mirror") at fixed $(r,\theta)$
around a black hole.
Let observer at $(r,\theta)$ with angular velocity $\Omega$ emit a flash
of light.
Some of the photons will get caught by the mirror and will skim along
its surface, circumnavigating the blackhole in the positive-$\phi$
direction.
Others will get caught and will skim along in the negative-$\phi$
direction.
Then, only the observer with his angular velocity
$\Omega=\omega(r,\theta)$ in Eq.~(\ref{eq:omega}) will receive back the photons from both directions simultaneously.
Only the observer regard the $+\phi$ and $-\phi$ directions as
equivalent in terms of local geometry.
Therefore, there is a preferred ``locally nonrotating" observer in this situation.
The metric inside the ring with respect to the locally nonrotating
observer takes the form:
\begin{eqnarray} \label{met:nonrot}
ds_{nr}^2 = - dt^2 + R^2 d\phi^2 ,
\end{eqnarray}
where we have ignored the $\theta$, $r$ coordinates and rescaled the
time coordinates since the observer are restricted to the ring.
The coefficient $R$ is independent of $t$ and $\phi$.
This metric is flat with respect to the coordinate $(t,R\phi)$.
However, there is no Lorentz boost-like symmetry which mixes $t$ and
$\phi$, since there is a preferred observer: {\it the locally
nonrotating observer}.
Therefore, the observer restricted to the ring can determine he is
rotating or not with respect to the locally nonrotating frame.
However, the locally nonrotating observer cannot determine whether he
is rotating or not with respect to an asymptotic infinity without
comparing his coordinates with respect to the asymptotic infinity.
Only after he comparing his coordinates with the asymptotic one, he can
determine he is rotating or not.

It is interesting to ask what takes away the apparent boost-like symmetry along $\phi$ in the metric~(\ref{met:nonrot}).
At first glance, one of the two is responsible for the breaking of the
symmetry, the work done by the mirror on the light and the compactness
of the angle $\phi$.
To examine these possibilities, we consider three limiting thought
experiments.
First, we take the vacuum ($M=J=0$) limit of the blackhole.
In this case, we have $\omega(r,\theta) =0$ and therefore, the locally
nonrotating frame selects the static frame.
The light bounces by the mirror to circulate the ring.
However, it should be noted that the force given by the mirror to the
light is orthogonal to the velocity of the light so that it does not
affect to the angular motion of light.
Second, we consider the $r\rightarrow \infty$ limit.
In this case, the experiment cannot select any observer since no light
can circumnavigate to return to the observer within finite time.
Therefore, there disappears the preferred locally nonrotating observer and the boost-like symmetry along $\phi$ will be restored.
Finally, let $r$ be placed at the last unstable circular orbit of photon (the photon sphere).
In this case, we do not need the mirror which restricts the path of the
light.
The observer may simply send light for both side of the $\phi$
directions and then waits until the light to arrive him after a full
circulation of the geodesic path.
There are no artificial work done by the mirror on the light, however, the light select the locally nonrotating observer.
This observation indicates that the breaking of the Lorentz boost-like
symmetry is not due to the work given by the mirror.
Since in this case, the light simply follows the geodesics, we may
accept the metric~(\ref{met:nonrot}) as a $2-$dimensional spacetime
metric with symmetry $\phi=\phi+2\pi$ without the $r$ and $\theta$
coordinates.
In conclusion, the breaking of the boost-like symmetry is solely
due to the compactness of the $\phi$ coordinate, not due to the work
given by the mirror.

Similarly, the boosted black string solution with the compact
$z$-coordinate is independent of the static one. Now consider
spacetime given by the metric~(\ref{Sol}) with the compact
$z$-coordinate with period $L$. We examine the frame dragging
effect by analyzing the metric from the point of view of moving
observer along the $z-$direction with velocity $q$ at
$r\rightarrow \infty$. The metric seen by the moving observer can
be obtained by using the remaining translational
symmetry\cite{Elvang}:
\begin{eqnarray} \label{coord}
z'= \gamma (z+q t),\quad \quad t'= \gamma^{-1} t ,
\end{eqnarray}
where $\gamma = \frac{1}{\sqrt{1-q^2}}$.
The metric~(\ref{Sol}),
from the point of view of moving observer with velocity $q$ at
$r\rightarrow \infty$, becomes
\begin{eqnarray} \label{met:compt}
ds^2= -\gamma^2 \left(F+2 q X-q^2H\right){dt'}^2 +2\left(X-q H
\right){dz'} {dt'} +\gamma^{-2} H {dz'}^2 .
\end{eqnarray}
The metric component $g_{tz'}$ is
\begin{eqnarray} \label{gtz}
g_{t'z'}&=& \left\{\begin{tabular}{ll}
 $\displaystyle -q+\frac{M\sinh2\xi (1 - q \tanh\xi)}{r }, $ & ~~~~Schwarzschild ,\\
 $\displaystyle -q-\frac{M\sinh2\xi (-q +\tanh\xi)}{r }, $ &~~~~ Kaluza-Klein bubble,\\
 \end{tabular} \right. \nn
\end{eqnarray}
whose asymptotic value is $-q$.
As a result, the metric~(\ref{met:compt}) with compact $z-$coordinate is
described by $3-$parameters $(M,\xi;q)$, where the set
$(M,\xi)$ denotes observer independent geometric properties and
$q$, the velocity of the observer.

The comoving velocity along $z$-coordinate at $r$ is
\begin{eqnarray} \label{eq:comov}
v(r) &=& -\frac{g_{t'z'}}{g_{z'z'}} = \gamma^2 \left(
q-\frac{X}{H} \right) ,
\end{eqnarray}
which asymptotically approaches to $q$ at $r\rightarrow\infty$  as it must.
For the boosted Schwarzschild solution the comoving velocity in Eq.~(\ref{eq:comov}) is
\begin{eqnarray} \label{eq:v:Schwarz}
v_{S}(r)=\gamma^2\left(  q -\frac{ \frac{M\sinh2\xi}{r}}{1+
\frac{2M\sinh^2\xi}{r}} \right) .
\end{eqnarray}
Note that the comoving velocity is dependent on $r$. Therefore, to
an observer moving with the velocity $q$, only the geometry at
$r_q=\frac{\sinh2\xi ~M}{q}(1-q \tanh\xi)$ is static since
$v_S(r_q)=0$ there. For different $r$, the geometry is stationary
and therefore there exists frame dragging. At the horizon the
velocity~(\ref{eq:v:Schwarz}) becomes $v_S(2M)= \gamma^2
(q-\tanh\xi)$, therefore the moving observers with velocity
$q=\tanh\xi$ see the horizon static, which is intuitively correct.

For the boosted Kaluza-Klein bubble solution, we have the comoving velocity,
\begin{eqnarray} \label{eq:v:bubble}
v_{KK}(r)= \gamma^2 \left( q + \frac{ \frac{M\sinh2\xi}{r}}{1-
\frac{2M\cosh^2\xi}{r}} \right).
\end{eqnarray}
To an observer moving with velocity $q$, only the geometry at
\begin{eqnarray} \label{rq}
r_q= 2M \cosh^2\xi(1-\frac{\tanh\xi}{q})
\end{eqnarray}
is static.
For different $r$, the geometry is stationary and therefore there exists frame dragging effect.
If an observer at infinity want to see a geometry at a given $r_q$ static,
he should move with the velocity $q$ satisfying Eq.~(\ref{rq}).
This velocity, however, becomes the light velocity at $r=2M\cosh^2\xi(1\pm \tanh \xi)$
and even larger than the light velocity in range between the two values of $r$.
Therefore, it is impossible that an observer at infinity see this region static.

At $r=2M$, the comoving velocity becomes $v_{KK}(2M)=\gamma^2
(q-\coth\xi)$. If we restrict the velocity $|q|< 1$, since the
motion of observer at infinity is restricted,
 we cannot make the velocity~(\ref{eq:v:bubble}) at $r=2M$ surface to zero.

The static limit is present at $r_e(q,\xi)$ satisfying
\begin{eqnarray} \label{static:2}
F+2 q X-q^2H =0 .
\end{eqnarray}
The static limit changes according to the velocity $q$ of the
observer at infinity. For the Schwarzschild case,
Eq.~(\ref{static:2}) gives
\begin{eqnarray*} \label{rhoe}
r_e &=& \frac{2M \cosh^2\xi(1 - q \tanh\xi)^2}{1-q^2} \geq 2M ,\nn
\end{eqnarray*}
where $r_e$ takes its minimum value $2M$ at $q=\tanh\xi$.
The ergo-region is located at $2M < r \leq r_e$.
To an observer moving with velocity $q=\tanh\xi$, there is no ergo-region since $r_e=2M$ overlaps with the horizon.
For the Kaluza-Klein bubble case, Eq.~(\ref{static:2}) gives
\begin{eqnarray*} \label{rhoe}
r_e &=& -\frac{2M \cosh^2\xi(q - \tanh\xi)^2}{1-q^2} .\nn
\end{eqnarray*}
Since we have restricted to $r\geq 2M$ for the bubble solution and the size of the velocity $q$ must be smaller than one,
 we have $r_e<0$ always. Therefore, there does not exist the ergo-region for the boosted bubble solution.

In discussion, the boosted black string solution in spacetime with
the compact $z-$coordinate is not equivalent to the static
solution and has a quantity $P$, the momentum with respect to an
asymptotically static observer, as a characteristic physical
parameter. As a result, the velocity frame dragging effect becomes
a physical observable.

The momentum parameter also becomes a physical observable for the
boosted Kaluza-Klein bubble solution which has an innate
compactified length. This compactness destroys the boost symmetry.
So, in the boosted Kaluza-Klein bubble solution, the momentum
parameter along compact direction becomes a physical quantity
which cannot be gauged away by boost transformation. In this
direction, it has been studied recently the thermodynamics of
boosted non-rotating black holes with the momentum along the
compact dimension~\cite{add}.

How does the velocity frame dragging effect appear from the point
of view of an observer in four dimensions?
The most distinguished fact is that there appears a spherical
ergo-region around the black string.
Therefore, if one find an object with spherically symmetric static
limit, it must be a black string solution with a compact
$z-$coordinate which has the velocity frame dragging effect.
This is different from the ergo-region around the Kerr blackhole,
which is not spherically symmetric.

It would be interesting to investigate the velocity frame dragging
effect for other black objects with compact dimensions and other gravity objects
such as stationary vacuum black string solution which is not equivalent to static solution \cite{Extra}.

\vspace{1cm}
\begin{acknowledgments}
We are very grateful to Gungwon Kang for helpful discussions. This
work was supported by the Daejin University Special Research Grants in 2007.
\end{acknowledgments} \vspace{1cm}


\begin{thebibliography}{10}

\bibitem{M}
K. Schwarzschild, Sitzber, Deut. Akad. Wiss. Berlin, K1. Math.-Phys.
Tech., s. 189 (1916).

\bibitem{Q1}
H. Reissner, Ann. Phys. {\bf 50} 106, (1916)

\bibitem{Q2}
G. Nordstr\"{o}m, Proc Kon, Ned. Acad. Wet. {\bf 20}, 1238,
(1918).

\bibitem{J}
R. P. Kerr, Phys. Rev. Lett. {\bf 11}, 237 (1963).

\bibitem{n}
E. T. Newman, E. Couch, K. Chinnapared, A. Exton, A. Prakash, and R.
Torrence, J. Math. Phys. {\bf 6}, 918 (1965).

\bibitem{nA_1}
P. B. Yasskin, Phys. Rev. D {\bf 12}, 2212 (1975).

\bibitem{nA_2}
R. Bartnik and J. Mckinnon, Phys. Rev. Lett. {\bf 61}, 141 (1988).

\bibitem{nA_3}
H. P. K\"{u}nzle, Commun. Math. Phys. {\bf 162}, 371 (1994).

\bibitem{nA_4}
B. Kleihaus, J. Kunz, and A. Sood (1997) [hep-th/9705179].

\bibitem{nA_5}
T. Tachizawa, K. Maeda, and T. Torii, Phys. Rev. D {\bf 51}, 4054 (1995).

\bibitem{dilatonic_1}
G. T. Horowitz, In Proceedings of {\it 1992 Trieste Spring School on String Theory and Quantum Gravity}, p.59 (1992) [hep-th/9210119].

\bibitem{dilatonic_2}
S. Hassan and A. Sen, Nucl. Phys. B {\bf 375}, 103 (1992).

\bibitem{dilatonic_3}
G. W. Gibbons and K. Maeda, Nucl. Phys. B {\bf 298}, 741 (1988).

\bibitem{dilatonic_4}
R. Kallosh, A. Linde, T. Ortin, A. Peet, and A. Van Proeyen, Phys. Rev. D {\bf 46}, 5278 (1992).

\bibitem{quantum_1}
B. A. Campbell, N. Kaloper, and K. A. Olive, Phys. Lett. B {\bf 285}, 199 (1992).

\bibitem{quantum_2} M. J. Bowick, S. B. Giddings, J. A. Harvey, G. T. Horowitz, and a. Strominger, Phys. Rev. Lett. {\bf 61}, 2823 (1988).

\bibitem{btz_1}
M. Ba\~{n}ados, C. Teitelboim, and J. Zanelli, Phys. Rev. Lett. {\bf 69}, 1849 (1992).

\bibitem{btz_2}
M. Ba\~{n}ados, C. Teitelboim, and J. Zanelli, Phys. Rev. Lett. {\bf 72}, 957 (1994).


\bibitem{LT_1}
H. Thirring, Physikalische Zeitschrift {\bf 19}, 33 (1918).

\bibitem{LT_2}
H. Thirring, Physikalische Zeitschrift {\bf 22}, 29 (1921).

\bibitem{LT_3}
J. Lense and H. Thirring, Physikalische Zeitschrift {\bf 19}, 156 (1918).

\bibitem{rot_1}
I. Ciufolini and E. C. Pavlis, Natue {\bf 431}, 958 (2004).

\bibitem{rot_2}
N. Ashby, Nature {\bf 431}, 918 (2004)

\bibitem{rot_3}
C. Seife, Science {\bf 306}, 502 (2004).

\bibitem{rot_4}
J. G. Williams et al. Phys. Rev. Lett. {\bf 93}, 261101 (2004).

\bibitem{rot_5}
I. H. Stairs et al., Phys. Rev. Lett. {\bf 93}, 141101 (2004).

\bibitem{chodos}
A. Chodos and S. Detweiler, General Relativity and Gravitation, {\bf 14}, 879 (1982).

\bibitem{hovdebo}
J. L. Hovdebo and R. C. Myers, Phys. Rev. D {\bf 73}, 084013 (2006).

\bibitem{inst_0}
R. Gregory and R. Laflamme, Phys. Rev. {\bf D 37}, 305 (1988).

\bibitem{inst_1}
R. Gregory and R. Laflamme, Phys. Rev. Lett. {\bf 70}, 2837 (1993).

\bibitem{inst_2}
S. S. Gubser and I. Mitra, JHEP {\bf 0108}, 018 (2001).

\bibitem{inst_3}
T. Hirayama and G. Kang, Phys. Rev. {\bf D 64}, 064010 (2001).

\bibitem{inst_4}
G. W. Kang and J. J. Lee, JHEP {\bf 0403}, 039 (2004), hep-th/0401225.

\bibitem{inst_5}
G. T. Horowitz and K. Maeda, Phys. Rev. Lett. {\bf 87}, 131301 (2001).

\bibitem{inst_6}
M. W. Choptuik, L. Lehner, I. I. Olabarrieta, R. Petryk, F. Pretorius, H. Villegas, Phys. Rev. {\bf D 68},  044001 (2003).


\bibitem{randal_1}
 L. Randall, R. Sundrum, Phys. Rev. Lett. {\bf 83}, 3370 (1999),
 hep-ph/9905221.

\bibitem{randal_2}
 L. Randall, R. Sundrum, Phys. Rev. Lett. {\bf 83}, 4690 (1999), hep-th/9906064.

\bibitem{hawking}
A. Chamblin, S. W. Hawking, and H. S. Reall, Phys. Rev. {\bf D61}, 065007 (2000).

\bibitem{harmark}
T. Harmark and N. Obers, hep-th/0503020.

\bibitem{misner}
C. W. Misner, K. S. Thorne, and J. A. Wheeler, "Gravitation, W. H. Freeman and company, New York, (1973)".

\bibitem{add}
D. Kastor, S. Ray, and J. Traschen, JHEP {\bf 0706} 026 (2007), [arXiv:0704.0729].

\bibitem{Elvang}
H. Elvang, R. Emparan, D. Mateos and H. S. Reall, JHEP {\bf 0508} 042 (2005).

\bibitem{Extra}
Hyeong-Chan Kim and Jungjai Lee, Phys. Rev. {\bf D 77}, 024012 (2008), [arXiv:0708.2469].

\end{thebibliography}

\vspace{4cm}

\end{document}